\documentclass[10pt,journal,compsoc]{IEEEtran}
\usepackage[nocompress]{cite}
\usepackage[pdftex]{graphicx}
\usepackage{amsmath}
\usepackage{algorithmic}
\usepackage{url}
\usepackage{soul}
\usepackage{colortbl}
\usepackage{collcell}
\usepackage[table,xcdraw]{xcolor}
\usepackage{multirow}

\usepackage{booktabs}
\newcolumntype{L}[1]{>{\raggedright\let\newline\\\arraybackslash\hspace{0pt}}m{#1}}
\newcolumntype{C}[1]{>{\centering\let\newline\\\arraybackslash\hspace{0pt}}m{#1}}
\newcolumntype{R}[1]{>{\raggedleft\let\newline\\\arraybackslash\hspace{0pt}}m{#1}}

\begin{document}
\title{Basic Ensembles of Vanilla-Style Deep Learning Models Improve Liver Segmentation From CT Images}

\author{A.~Emre~Kavur,~\IEEEmembership{Member,~IEEE,}
        Ludmila~I.~Kuncheva, 
        and~M.~Alper~Selver,~\IEEEmembership{Member,~IEEE}%
\IEEEcompsocitemizethanks{\IEEEcompsocthanksitem A.E. Kavur is with the Graduate School of Natural and Applied Sciences, Dokuz Eylul University, Izmir, TURKEY e-mail: emrekavur@gmail.com.
\IEEEcompsocthanksitem L. Kuncheva is with the School of Computer Science and Electronic Engineering, Bangor University, Bangor, UNITED KINGDOM  e-mail: l.i.kuncheva@bangor.ac.uk. 
\IEEEcompsocthanksitem M.A. Selver is with the Department of Electrical and Electronics Engineering, Dokuz Eylul University, Izmir, TURKEY e-mail: alper.selver@deu.edu.tr }
}

\IEEEtitleabstractindextext{%
\begin{abstract}
Segmentation of the liver from 3D computer tomography (CT) images is one of the most frequently performed operations in medical image analysis. In the past decade, Deep Learning Models (DMs) have offered significant improvements over previous methods for liver segmentation. The success of DMs is usually owed to the user's expertise in deep learning as well as to intricate training procedures. The need for bespoke expertise limits the reproducibility of empirical studies involving DMs. Today's consensus is that an ensemble of DMs works better than the individual component DMs. In this study we set off to explore the potential of ensembles of publicly available, `vanilla-style' DM segmenters Our ensembles were created from four off-the-shelf DMs: U-Net, Deepmedic, V-Net, and Dense V-Networks. To prevent further overfitting and to keep the overall model simple, we use basic non-trainable ensemble combiners: majority vote, average, product and min/max. Our results with two publicly available data sets (CHAOS and 3Dircadb1) demonstrate that ensembles are significantly better than the individual segmenters on four widely used metrics.
\end{abstract}

}

\maketitle
\IEEEdisplaynontitleabstractindextext
\IEEEpeerreviewmaketitle

\section{Introduction}

\IEEEPARstart{S}{egmentation} has been recently reported to be the most studied field of biomedical image processing, accounting for around 70\% of all studies~\cite{Maier-Hein2018}. Liver segmentation is a particularly important topic in medical image segmentation due to its use in numerous clinical procedures including but not limited to volumetry~\cite{Lu2017}, early-stage diagnosis~\cite{Moghbel2018}, tumour/lesion detection~\cite{Li2018,Chlebus2018,Christ2016,Vorontsov2019}, disease classification~\cite{Bal2018}, surgery, and radiotherapy treatment planning~\cite{YANG201841}. Among many emerging machine learning paradigms, Deep Learning Models (DMs) and particularly Convolutional Neural Networks (CNNs), have achieved remarkable results~\cite{ee5ac00b4eb14ad4987ed75a0664a371}. The two main requirements of DMs are 1) the availability of extensively annotated datasets for training, and 2) user experience and expertise for constructing successful architectures and adjusting their parameters. The vibrant interest in the area of medical image segmentation has prompted the organisation of contests and challenges, whereby common databases are released. Over the last five years, in a typical medical imaging challenge, participants prefer to design a dedicated DM or implement and tune a previously designed DM such as U-Net, Deepmedic, V-Net, Dense V-Networks, etc. \cite{Kamnitsas2016, Ronneberger2015, Milletari2016, Gibson2018a}. 

Available 3D volumetric data sets contain only a few tens of images~\cite{Heimann2009,Bilic2019,Menze2015} due to the high expense of gathering and annotating such data sets. This number is far too small for proper training of a DM, and could lead to spurious results due to overfitting. 
Classifier ensembles are known to achieve better results compared to their base classifiers~\cite{Kuncheva2014} even when those classifiers are overtrained. Organisers of medical segmentation competitions often demonstrate the ensemble superiority combining the top 5 or so entries in the league table.~\cite{Kamnitsas2017b, Isensee2019}
The practical use of such ensembles is questionable because of the following. Multiple submissions are normally allowed, whose results are disclosed to the participant. This allows the participant to tune their model on the {\em testing data}, which is a form of `peeking'~\cite{Smialowski10,Reunanen03,Diciotti13}. Therefore, an ensemble of contest winners may be over-tuned on the testing data. Besides, the ensemble members are likely to be sophisticated (and not always reproducible) models of their own. 

Contrary to most recent studies, we propose, what we call, a `vanilla'-style ensemble where the DMs are state-of-the-art baseline models with no change of their default parameters. Next, we propose four basic combination methods as the ensemble combiners for the following reason. Given the small size of the training data (number of patients),  further tuning is not advisable~\cite{Duin02}. The vanilla ensemble gives the practitioner a ready-made solution, eliminating the need for elaborate tuning and structure modifications of the model, both of which require bespoke expertise and sometimes just luck.  

The rest of the paper is organised as follows. Section~\ref{sec:relatedw} summarises the related work on medical image segmentation using DMs and ensembles thereof. Section~\ref{sec:methods} gives details of the four DMs used in this study, the ensemble combination rules, and the four metrics used to evaluate the segmentation results. Our motivation for using ensembles is illustrated through an example.
The experiment with the CHAOS competition data set~\cite{chaosdata} and the 3Dircadb dataset~\cite{3Dircadb} is reported in Section~\ref{sec:expe}. Finally, Section~\ref{sec:concl} gives our conclusions and outlines future research directions.

\section{Related work}
\label{sec:relatedw}

Deep learning has rightfully attracted widespread attention in the literature by outperforming alternative machine learning approaches in various fields of applications, especially in medical imaging~\cite{Greenspan16,Shin16}. 

Further on, combining multiple results coming from different models to achieve a final refined outcome has become an effective way to obtain superior results~\cite{Kuncheva2014}. It has been recognised that most high-profile competitions such as Imagenet\footnote{\url{http://www.image-net.org/challenges/LSVRC/}} and Kaggle\footnote{\url{https://www.kaggle.com/competitions}} are won by ensembles of deep learning architectures~\cite{Huang17}. Lately, ensembles of DMs have been proposed for various domain applications, for example, object detection~\cite{Raznikov19}, video classification~\cite{Zheng19}, aerial scene classification~\cite{Dede19}, and diagnosis and prediction in industrial systems and processes~\cite{Ma19,Zhang17}. 

Usually, a small group of DMs are taken as the ensemble members due to high computational cost. The models are trained either separately~\cite{Kamnitsas2017,Warfield04} or simultaneously, together with another network which combines them~\cite{Dede19,Ma19,Zheng19}. Some of the ensemble methods train {\em different} DMs (heterogeneous ensembles) while others train the same DM with different parameters, training pattern or data. An interesting training strategy which avoids training multiple models is the ``snapshot ensembling''~\cite{Huang17}, whereby the training process of a single DM is stopped at different local minima, and the respective DM is retrieved. The ensemble is then constructed from these (possibly undertrained but supposedly diverse) DMs. 

For individually trained DMs, simple ensemble combination rules are applied such as majority vote~\cite{Ortiz2016}, average~\cite{Kamnitsas2017b,MajiSMS16,Codella17}, product and more~\cite{Warfield04,Ju18}. Warfield et al.~\cite{Warfield04} propose an expectation maximisation method named STAPLE for the combination of image segmenters.\footnote{\url{http://crl.med.harvard.edu/software/}} Trained combination rules such as stacked generalisation, Bayes models and ``super learner''~\cite{Ju18} have also been considered for this task. 

The remarkable potential of ensembles of DMs in medical imaging is often illustrated at the end of public competitions~\cite{Prevedello19}. The top methods (usually DMs) are combined through majority voting, and the result is usually better than the best contestant's result~\cite{Menze2015,Toro2016,Bilic2019,Kamnitsas2017b,Kavur2019a}. While this makes a compelling case for ensembles of DMs, such results could be misleading. The individual ensemble members (the contestant entries) have been honed on the testing data during the competition, which means that the testing data is no longer independent on the training data, and the result of the ensemble (and the individual members) maybe optimistically biased. (The ``peeking'' phenomenon.)

A multitude of DM ensembles has been proposed specifically for medical image segmentation~\cite{Ju18,Codella17}.
Bilic et al.~\cite{Bilic2019} published a compelling overview of the results of the Liver Tumour Segmentation Benchmark (LiTS) contest held in 2017.\footnote{Organised in conjunction with the IEEE International Symposium on Biomedical Imaging (ISBI) 2017 and the International Conference On Medical Image Computing \& Computer Assisted Intervention (MICCAI) 2017.} The study analyses the winning state-of-the-art models in the competition. Although the focus is on segmenting tumours in the liver, the segmentation of the liver itself is also considered. Among other conclusions, the authors note the following:
\begin{enumerate}
    \item It is still difficult to provide recommendations with regards to the exact network design (structure, parameters, modifications, training). The current solutions are mostly guided by rough ideas instead of strict, proven guidelines. Exploring the possible choices for each task at hand is largely hindered by long training times.
    \item Only a few of the best DMs were fully-3D (that is, taking as input  a volume image). This was again attributed to excessive computational cost. The authors reported that potentially, full 3D DMs would be more successful than the currently used sparse 3D or 2.5D DMs. Here we can add a comment that full 3D DMs may need even more parameters to be set up compared to the non-3D models, which makes the practitioner's task harder.
    \item It was observed that ensemble methods outperformed in general the single segmenter methods but their practical use may be hampered by computational constraints. In other words, it is questionable whether the gain the ensembles offer justifies the extra computational resources needed. Our experiments here show that ensembles are significantly better than their individual components, and therefore we argue that the extra cost is well justified. 
\end{enumerate}

To address the first concern, here we propose to use off-the-shelf (vanilla-style) DMs. This will eliminate the need to tune any parameters, modify the structure or devise a bespoke training protocol. Second, we propose to use full 3D DMs, which are available at this stage, and their training takes a reasonable time. Third, we will demonstrate that the individual segmenters are outperformed by the ensemble segmenters by a large margin, and therefore we advocate the ensemble approach.

We chose to use a heterogeneous ensemble of four 3D DMs because this ensemble type was found to be superior to homogeneous ensembles of DMs, even to those trained by snapshot ensembling~\cite{Dede19}. We kept the combination rules as simple as possible for two reasons. First, we eliminate further overfitting of the ensemble, and second, simple non-trainable combiners are straightforward to implement and do not require profound expertise in either DMs or ensemble methods.  

\section{Methods}
\label{sec:methods}
\subsection{DMs for liver segmentation}
\label{sec:dms}
The four well-established CNNs used as our ensemble members are detailed below.  

\subsubsection{U-Net}
U-Net is one of the first convolutional neural networks designed for the segmentation of biomedical images.\cite{Long2015}.
U-Net has been designed to operate with fewer training data compared to standard CNNs without compromising the segmentation accuracy. 
In this work, we chose the original implementation of U-Net from NiftyNet  with cross-entropy as the loss function \footnote{\url{https://niftynet.readthedocs.io/en/dev/_modules/niftynet/network/unet.html}}. 
NiftyNet provides an open-source front-end platform for different CNN solutions \cite{Gibson2018, Li2017} for the assessment of medical images. NiftyNet offers modular design so that different CNNs can be construcetd.

\subsubsection{Deepmedic}
DeepMedic is a multi-scaled 3D Deep Convolutional Neural Network combined with a linked 3D fully connected Random Field~\cite{Kamnitsas2016}. Deepmedic was originally designed for brain lesion segmentation and won ISLES 2015 and BraTS 2017 challenges~\cite{Kamnitsas2017}. 
Deepmedic was used from its original source. The system was directly downloaded and applied to our data. \footnote{\url{https://github.com/deepmedic/deepmedic}}

\subsubsection{V-Net}
V-Net is designed for volumetric segmentation of the prostate from MR image series \cite{Milletari2016}. The whole pathway has a V-shape, which is where the CNN gets its name. While V-Net and U-Net share similar structures, their loss functions are different. 
The V-Net model used here was also sourced from NiftyNet. \footnote{\url{https://niftynet.readthedocs.io/en/dev/_modules/niftynet/network/vnet.html}} 

\subsubsection{Dense V-Networks}
Dense V-Networks are developed for automatic segmentation of abdominal organs from CT scans~\cite{Gibson2018a}. They differ from the other DMs by including three dense feature blocks at each encoding stage. They use the probabilistic Dice score as their loss function. 

The Dense V-network model used here was taken from NiftyNet. \footnote{\url{https://niftynet.readthedocs.io/en/dev/_modules/niftynet/network/dense_vnet.html}} 

\subsection{Ensemble combination methods}
\label{sec:combiners}
We consider the simplest ensemble combination methods which do not require any further training or parameter tuning~\cite{Kuncheva2014}. Denote by $p_1,\ldots, p_4$ the values of the probability maps outputted by the four segmenters for a given voxel in the 3D image. These values estimate the probability that the voxel is from the class foreground. The combination methods in this paper are as follows: 

\medskip\noindent
$\bullet $
{ \em Majority Voting.} The probability map of each segmenter is thresholded at 0.5. The voxels with values above this threshold are considered foreground (liver) and the others, background. Having four segmenters, each voxel receives four votes. The voxel is labelled as foreground by the Majority Voting method if 3 or 4 votes are for the foreground. 

\medskip\noindent
$\bullet $
{\em Average combiner.} For each voxel, we calculate the support for class foreground by averaging the four probability map values:
$P_{\tiny foreground} = \frac{1}{4}(p_1+p_2+p_3+p_4).$ If $P_f>0.5$, the class foreground is assigned to that voxel. The class background is assigned otherwise.
    
\medskip\noindent
$\bullet $
{\em Product combiner.} For each voxel, we calculate the support for class foreground by $P_f=-\log(P_{f0}) + \sum_1^4{\log(p_i)}$,
where any base logarithm can be used, and $P_{f0}$ denotes the prior probability of class foreground. This probability can be estimated from the training data as the proportion of foreground voxels in all images. We next calculate the support for class background by $P_b=-\log(1-P_{f0}) + \sum_1^4{1-\log(p_i)}$.
We assign class foreground to the voxel if $P_f>P_b$, and class background, otherwise.

\medskip\noindent
$\bullet $
{\em Min/Max combiner.} The minimum and the maximum combination rules are identical for the case of two classes~\cite{Kuncheva2014}, which is why we report one value for both. In this combiner, we calculate $P_f=\min_i(p_i)$ and $P_b=\min_i(1-p_i)$.
Again, we assign class foreground to the voxel if $P_f>P_b$, and class background, otherwise.

\subsection{Evaluation Metrics}
The primary aim of medical image segmentation is to develop tools for clinical needs such as diagnosing, surgery planning, and organ transplant operations. Hence, tolerance of error is minimal. According to previous studies~\cite{Maier-Hein2018,Yeghiazaryan2015}, there is no single metric that evaluates 3D segmented data completely and fairly in terms of clinically acceptable results. Since the results have to be analysed from many perspectives, the aggregation of multiple evaluation metrics was preferred \cite{Langville2013}.  Spatial overlap-based, volume-based, and spatial distance-based metrics were chosen here to analyse different aspects of segmentation in terms of different aspects. 

The four metrics used to evaluate performance for a segmentation result are listed below:
\begin{enumerate}
\item Dice coefficient (DICE). Denoting the set of foreground voxels in the candidate segmentation by $X$ and that for the ground truth by $Y$, the Dice coefficient is calculated as Dice $= 2|X\cap Y|/(|X|+|Y|)$, where $|.|$ denotes cardinality (the larger, the better).
\item Relative absolute volume difference (RAVD). Using the notation above, RAVD = $\rm{abs}(|X|-|Y|)/|Y|$, where `abs' denotes the absolute value (the smaller, the better). 
\item Average symmetric  surface distance (ASSD). This metric is the average Hausdorff distance (in millimetre) between border voxels in $X$ and $Y$ (the smaller, the better).  
\item Maximum symmetric  surface distance (MSSD). This metric is the maximum Hausdorff distance (in millimetre) between border voxels in $X$ and $Y$ (the smaller, the better).
\end{enumerate}

The code for all metrics (in MATLAB, Python and Julia) are available at \textit{\url{https://github.com/emrekavur/CHAOS-evaluation}}, where we also provide the metrics' calculation details and CHAOS scoring system.

\subsection{Why are ensembles better?}

Ensembles of {\em diverse} segmenters can clear erroneous artefacts and smooth out spurious contours in the individual segmentations. This is expected to lead to a better overall match to the ground truth.

\begin{figure}[htb]
\centering
\begin{tabular}{cc}
U-net &
Deepmedic\\
\includegraphics[width = 0.22\textwidth]{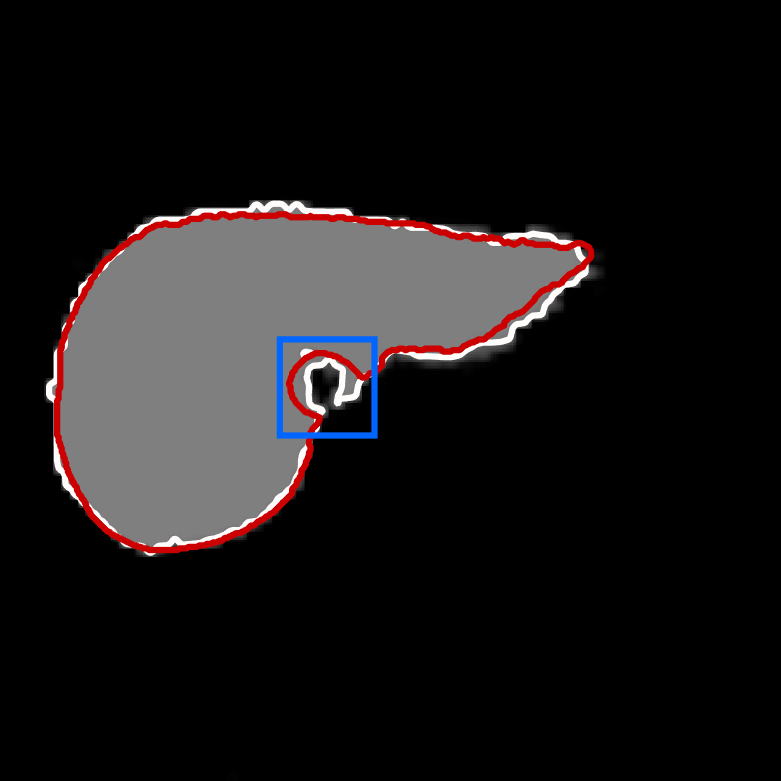}&
\includegraphics[width = 0.22\textwidth]{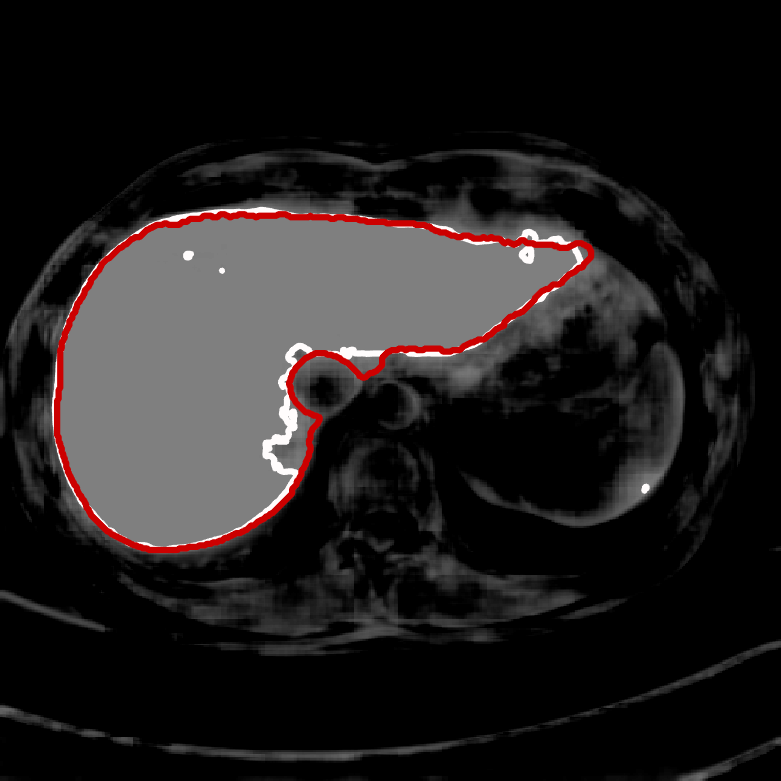}\\
&\\
V-net &
Dense V-networks\\
\includegraphics[width = 0.22\textwidth]{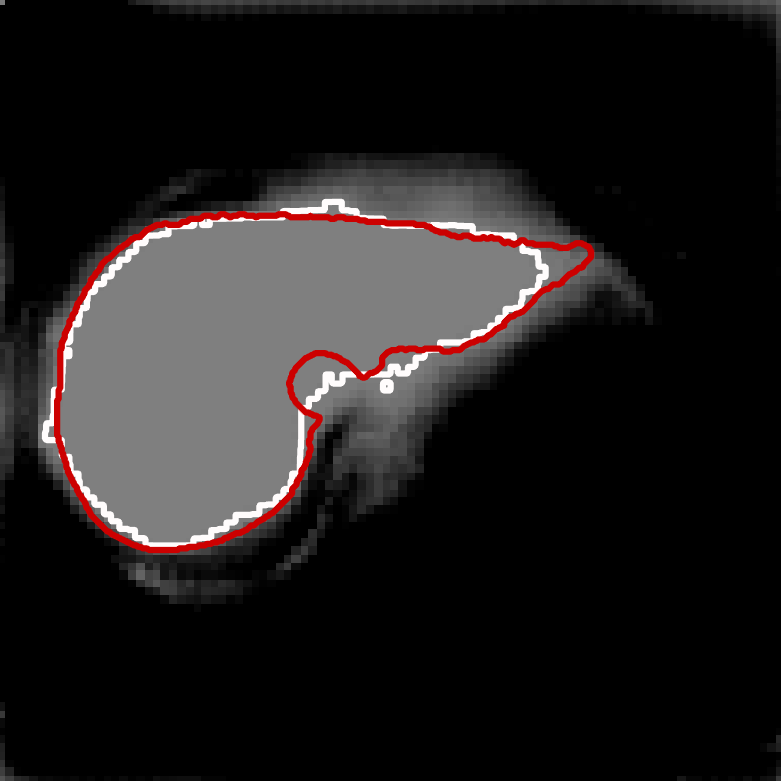}&
\includegraphics[width = 0.22\textwidth]{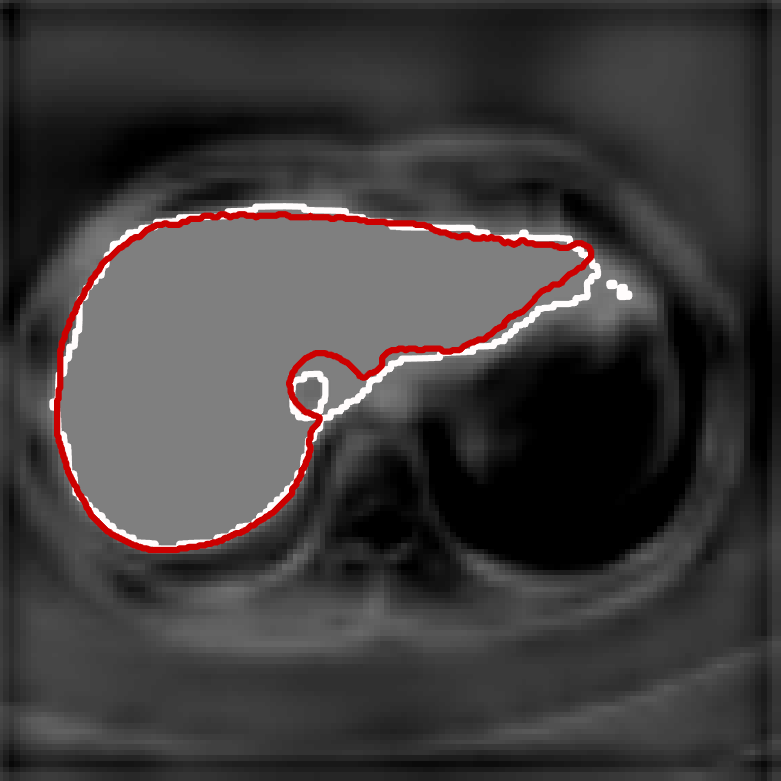}\\

\end{tabular}

\caption{Illustration of the segmentation results of the individual segmenters. White lines represent the border of the segmentation results while red lines represent the border of the ground truth.}
\label{liver_ind_results}
\end{figure}

Figure~\ref{liver_ind_results}  shows the results of the four individual segmenters. The grey level intensity reflects the probability map of foreground versus background. The white contour is the proposed segmentation boundary obtained by thresholding the probability map at 0.5. The red contour is the ground truth. \footnote{The image for this example is slide 11 from the data of patient number 4 in the CHAOS dataset.}

\begin{figure*}[hbt]
\centering
\begin{tabular}{ccccc}
Ensemble& U-net & Deepmedic& V-Net & Dense V-Net\\
\includegraphics[width = 0.18\textwidth]{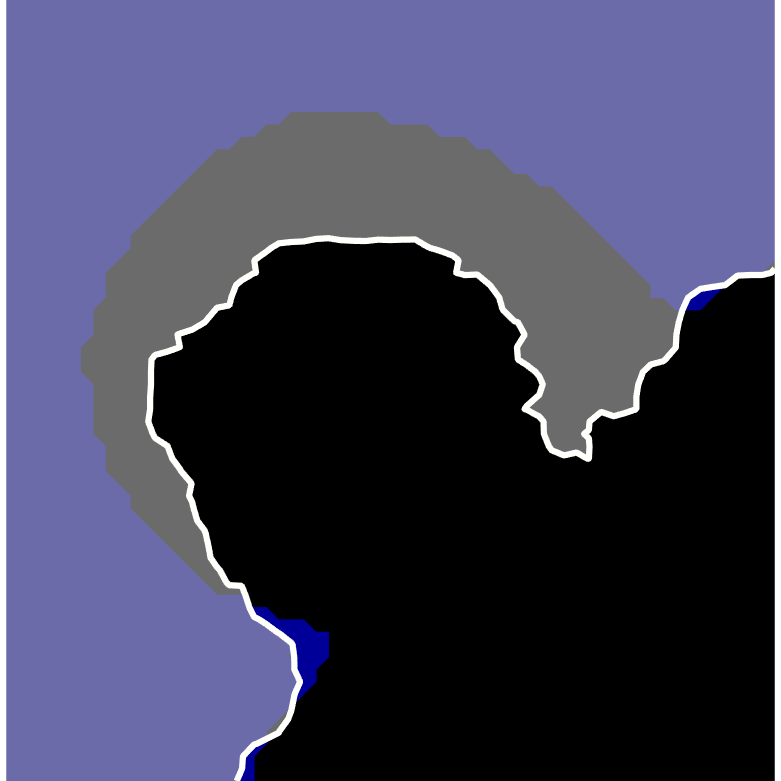}&
\includegraphics[width = 0.18\textwidth]{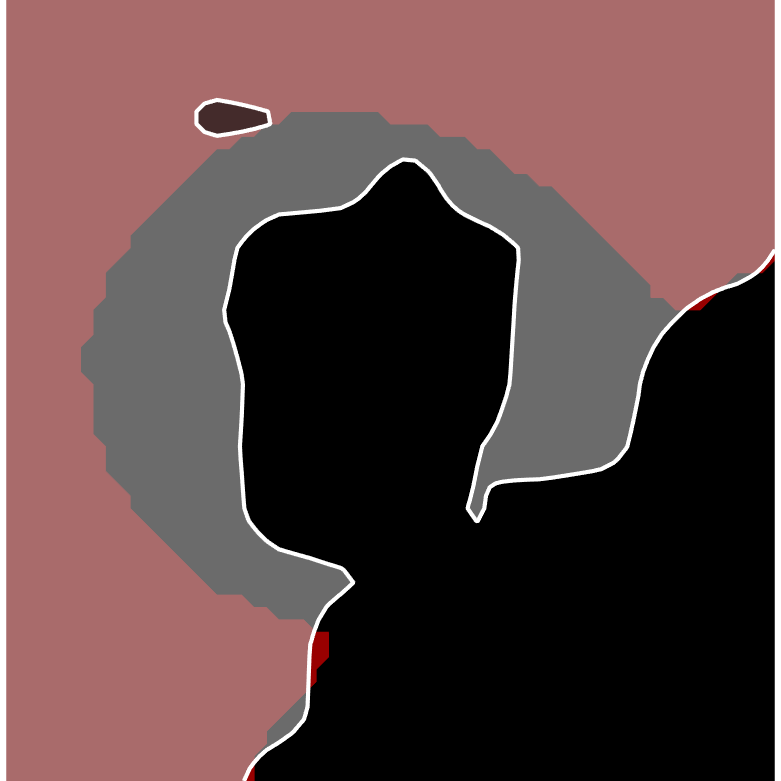}&
\includegraphics[width = 0.18\textwidth]{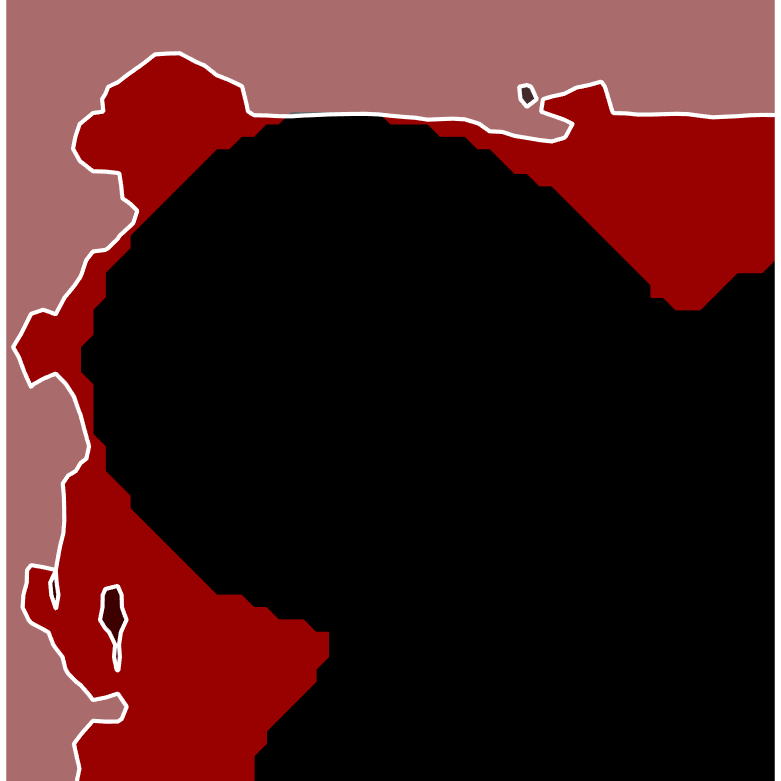}&
\includegraphics[width = 0.18\textwidth]{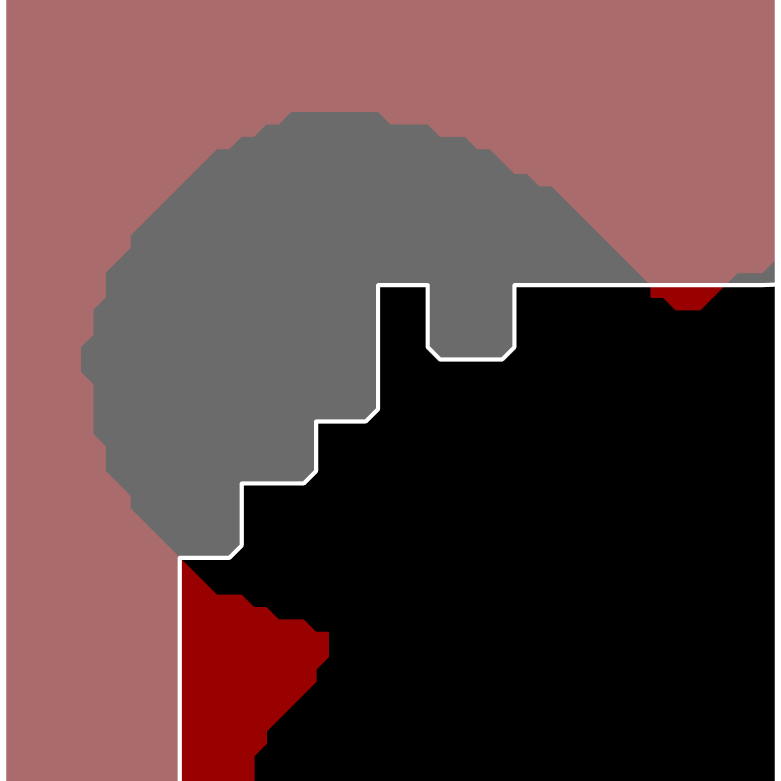}&
\includegraphics[width = 0.18\textwidth]{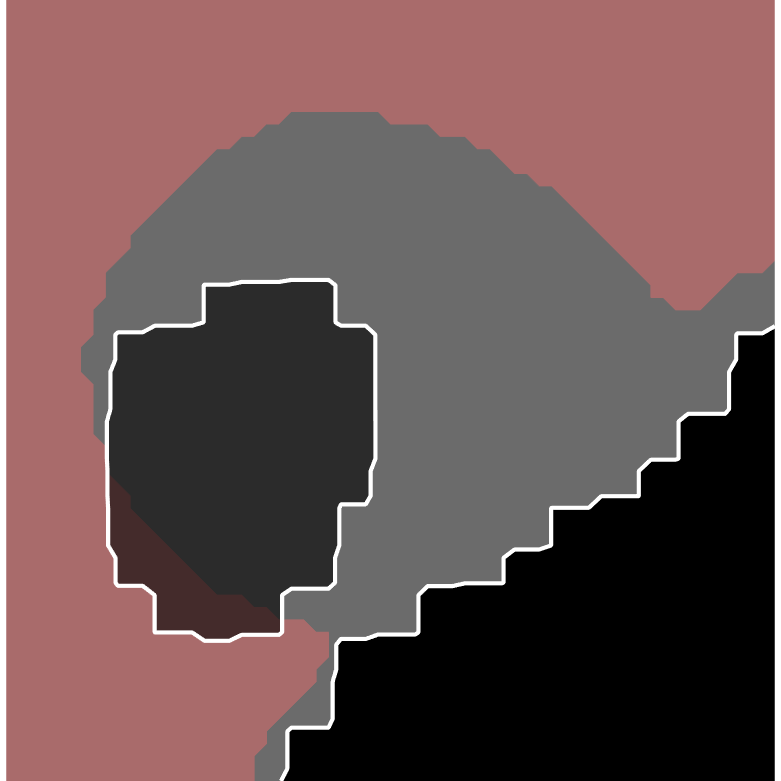}\\

0.7306&    0.7003&    0.5265 &0.6175&  0.5863\\
\end{tabular}

\caption{The ensemble segmentation with the product combination rule and the individual segmentations zoomed on {\it vena cava superior} (the blue rectangle in the U-net plot in Figure~\ref{liver_ind_results}). The ground truth is shaded in blue in the ensemble plot, and in red in the plots for the individual segmenters. The guessed segmentation is overlaid in transparent grey. The Dice score for the chosen area is shown under each plot.} 
\label{liver_map_results}
\end{figure*}

In order to show the advantage of the ensemble, we chose a small, notoriously difficult, region to zoom on: {\it vena cava superior} (the blue rectangle in the U-Net plot in Figure~\ref{liver_ind_results}). Figure~\ref{liver_map_results} contains five plots of the segmented region. The product combiner was chosen for the ensemble.  The ground truth is shaded in blue in the ensemble plot, and in red in the plots for the individual segmenters. The guessed segmentation is overlaid in transparent grey. The Dice score for the chosen area is shown under each plot. A Dice score of 1 indicates perfect segmentation while lower values indicate mismatch. As the results show, both visually and through the numbers, the ensemble segmentation is better than any of the individual ones.

\section{Experimental evaluation}
\label{sec:expe}

Of course, one example can serve as an illustration but is not proof. To give empirical support to our claim that vanilla-style ensembles of DMs are superior to the individual DMs, we carry out an experiment with two publicly available datasets.

The purpose of this experiment is twofold. First, we demonstrate the overfitting effect exhibited by DMs on a relatively small data set. The second branch of our experiment explores the four classifier ensemble combiners for liver segmentation from CT scans. Implementation and evaluation of ensemble methods codes for a sample patient CT set are available at \textit{\url{https://github.com/emrekavur/Basic-Ensembles-of-DMs-for-Liver-Segmentation}} page.

\subsection{Data}
In order to demonstrate the consistency of the results, we performed our experiments on two publicly available databases that were published at different times and have different characteristics.\\

\noindent $\bullet$ {\em CHAOS Competition Data}

\medskip
The data consist of 40 sets of images taken from the Combined (CT-MR) Healthy Abdominal Organ Segmentation Challenge (CHAOS) \cite{chaos, chaosdata}. CHAOS is a medical imaging challenge focused on segmentation of healthy abdominal organs from CT and/or MRI image series~\cite{kavur2020chaos}. CHAOS was started during the IEEE International Symposium on Biomedical Imaging (ISBI) on April 11, 2019, Venice, Italy. CHAOS has five independent tasks related to different modalities and abdomen organs. Participants may join the challenge for a single task or for multiple tasks.\footnote{ As of January 2020, CHAOS is one of the most popular medical imaging challenges in \url{grand-challenge.org} and has 1516 current participants at the  website~\cite{chaos, kavur2020chaos}}.

In this work, the CT part of the CHAOS data was used. The CT data contains abdomen images of 40 different patients who have healthy livers. The technical information about the data is presented in Table~\ref{tab:data_stats}.


\begin{table}[b]
\centering
\def\arraystretch{1.3}
\caption{Statistics about CHAOS CT and 3DIRCADb datasets.}
\label{tab:data_stats}
\begin{tabular}{L{4.74cm}C{1.5cm}C{1.5cm}}
\toprule
 & 3DIRCADb & CHAOS\\
\midrule
Number of 3D image sets & 20 (10 + 10) & 40 (20 + 20) \\ 
Spatial resolution of files & 512 x 512 & 512 x 512   \\ 
Number of files (slices) in all cases [min--max] & [74 -- 260] & [78 -- 294] \\ 
Average number of files in the cases & 141 & 160 \\ 
Total number of files in the dataset & 2823 & 6407 \\ 
X space (mm) [min--max] & [0.56 -- 0.87] & [0.54 -- 0.79] \\ 
Y space (mm) [min--max] & [0.56 -- 0.87] & [0.54 -- 0.79] \\ 
Slice thickness (mm) [min--max] & [1.60 -- 4.00] & [2.00 -- 3.20] \\
\bottomrule
\end{tabular}
\end{table}

The training part of the data contains 20 randomly chosen anonymised DICOM images and their ground truths. However, the test data includes only the anonymised DICOM images of the remaining 20 patients. The evaluation of the results on the test data is done by an evaluation script hosted on the \url{grand-challenge.org} website. This ensures that CHAOS competitors do not have access to the ground truth (manual segmentation) of the testing data. However, since multiple submissions are allowed and the overall score across the testing data is returned to the participant, the competitors can gauge the success of their solution and tune it accordingly (the peeking phenomenon). In our experiments, we report results on the testing data, but we do not tune any of our methods on these results.

\bigskip
\noindent $\bullet$ {\em 3Dircadb1 Data}
\medskip

The second dataset, 3D-IRCADb-01 (3D Image Reconstruction for Comparison of Algorithm Database) \cite{3Dircadb} contains abdomen CT scans of 20 patients. Unlike CHAOS dataset, 75\% of cases have hepatic tumours in the liver. 3D-IRCADb not only provides ground truths of livers but also various structures such as all hepatic veins and hepatic tumours. The ground truths were created by clinical experts. Except for tumours, all structures inside of the liver were considered together as segmentation target in this work. We equally divided 3D-IRCADb into two parts for training and testing (10 sets for each). The technical information about each dataset is presented in Table~\ref{tab:data_stats}. With 3Dircadb1 data, the same procedures were followed with CHAOS data for developing, training, and testing.

\subsection{Experimental protocol}
For each dataset, the four DMs were trained on the training data. For an input image (3D CT scan), each DM returns a 3D probability map $P$. For voxel at $(i,j,k)$, $P(i,j,k)$ is the probability that the voxel belongs to the liver. The output probability maps are then smoothed using 
$3\times3\times3$ convolution kernel (function \verb|smooth3| in MATLAB) to eliminate small defects in the segmentation. We also smoothed in the same way the ensemble output after applying the respective combination rule.

Next, we evaluated the four metrics on the training data and then on the testing data, both for the individual DMs and the four ensembles. 

Finally, we tested the hypothesis that ensembles are better than individual DM segmenters on the testing data by running tests for each (ensemble, segmenter) pair, for each of the four metrics.

To determine the statistical significance of the difference between methods $A$ and $B$, we applied the following protocol:
\begin{itemize}
    \item {\em Paired samples.} Suppose that $x$ and $y$ are the vectors containing the values for methods $A$ and $B$, respectively. (For example, $x$ may contain the Dice scores obtained from Deepmedic for the 20 testing patients in the CHAOS data, and $y$ may contain the Dice scores obtained from the Majority vote ensemble for the same 20 patients.) We are interested whether the means of $x$ and $y$ are significantly different. Using Lilliefors goodness-of-fit test, we check the normality of the difference $x-y$. If normality cannot be rejected at 0.05 level, we use the paired t-test for comparing the means. If the normality of the difference does not hold, we use Wilcoxon signed-rank test for zero median of the difference.
    
    \item {\em Non-paired samples.} For non-paired samples, we check normality of $x$ and $y$. If both hold, we apply 2-sample t-test. Otherwise, we use the Wilcoxon rank-sum test (Mann–Whitney U test).   
\end{itemize}

\subsection{Results}

A full set of results (all metrics for all DMs and all ensembles; training and testing) is provided in \textit{\url{https://github.com/emrekavur/Basic-Ensembles-of-DMs-for-Liver-Segmentation}} 
page. Tables~\ref{tab:res1} -- \ref{tab:res4} show the average results for the two datasets, all metrics, individual DMs and ensembles. 

\begin{table*}[ht]
\begin{minipage}{0.49\textwidth}
\centering
\caption{Metric results on CHAOS training data for the individual segmenters and the ensemble methods. The circle marker indicates results where the overfitting was not found to be significant.}\vspace{-2mm}
\label{tab:res1}
\centering
\def\arraystretch{1.2}
\begin{tabular}{l|rrrr}
& DICE & RAVD & ASSD & MSSD \\ \hline
U-Net& 0.935& $\circ$14.800& 3.903& 54.650\\
Deepmedic& 0.984& 1.115& 1.709& 67.078\\
V-Net& 0.948& $\circ$3.824& 1.656& 42.972\\
Dense V-networks& 0.932& 3.039& 2.289& $\circ$78.118\\
\hline
\rowcolor[rgb]{0.9 0.9 0.9}
Average& 0.950& 5.694& 2.389& 60.705\\
\hline
\rowcolor[rgb]{0.8 0.8 1}
Majority Vote& 0.976& 2.401& 0.746& 11.043\\
\rowcolor[rgb]{0.8 0.8 1}
Average& 0.981& 1.003& 0.637& 11.621\\
\rowcolor[rgb]{0.8 0.8 1}
Product& 0.975& $\circ$3.493& 0.888& 12.581\\
\rowcolor[rgb]{0.8 0.8 1}
Min-Max& 0.978& 1.208& 0.811& 11.559\\
\end{tabular}
\vspace{2mm}
\end{minipage}%
\hspace{5mm}
\begin{minipage}{0.49\textwidth}
\centering
\caption{Metric results on CHAOS test data for the individual segmenters and the ensemble methods. The best value in each column is enclosed in a box.}\vspace{-2mm}
\label{tab:res2}
\centering
\def\arraystretch{1.2}
\begin{tabular}{l|cccc}
& DICE & RAVD & ASSD & MSSD \\ \hline
U-Net& 0.811& 54.842& 14.253& 104.515\\
Deepmedic& 0.951& \fbox{3.058}& 7.174& 141.473\\
V-Net& 0.879& 17.434& 6.146& 104.189\\
Dense V-networks& 0.886& 7.702& 4.492& 113.139\\
\hline
\rowcolor[rgb]{0.9 0.9 0.9}
Average& 0.882& 20.759& 8.016& 115.829\\
\hline
\rowcolor[rgb]{1 0.8 0.8}
Majority Vote& 0.952& 4.235& \fbox{1.719}& \fbox{28.517}\\
\rowcolor[rgb]{1 0.8 0.8}
Average& \fbox{0.953}& 3.839& 1.956& 30.676\\
\rowcolor[rgb]{1 0.8 0.8}
Product& 0.946& 6.867& 2.121& 32.696\\
\rowcolor[rgb]{1 0.8 0.8}
Min-Max& 0.937& 6.094& 2.311& 35.052\\
\end{tabular}
\vspace{2mm}
\end{minipage}
\hspace{0.10\textwidth}
\begin{minipage}{.49\textwidth}
\centering
\caption{Metric results on 3Dircadb1 training data for the individual segmenters and the ensemble methods. The circle marker indicates results where the overfitting was not found to be significant.}\vspace{-2mm}
\label{tab:res3}
\centering
\def\arraystretch{1.2}
\begin{tabular}{l|rrrr}
& DICE & RAVD & ASSD & MSSD \\ \hline
U-Net& 0.904& 19.320& $\circ$7.075& $\circ$70.798\\
Deepmedic& 0.988& 0.216& 0.399& 36.127\\
V-Net& 0.968& 2.974& 0.975& 17.587\\
Dense V-networks& 0.973& 1.110& 0.930& $\circ$51.780\\
\hline
\rowcolor[rgb]{0.9 0.9 0.9}
Average& 0.958& 5.905& 2.345& 44.073\\
\hline
\rowcolor[rgb]{0.8 0.8 1}
Majority Vote& 0.979& 2.432& 0.633& 12.913\\
\rowcolor[rgb]{0.8 0.8 1}
Average& 0.982& 1.844& 0.615& 18.759\\
\rowcolor[rgb]{0.8 0.8 1}
Product& 0.978& $\circ$3.426& 0.841& 20.764\\
\rowcolor[rgb]{0.8 0.8 1}
Min-Max& 0.980& 2.052& 0.712& 19.423\\
\end{tabular}
\end{minipage}%
\hspace{5mm}
\begin{minipage}{0.49\textwidth}
\centering
\caption{Metric results on 3Dircadb1 test data for the individual segmenters and the ensemble methods. The best value in each column is enclosed in a box.}\vspace{-2mm}
\label{tab:res4}
\centering
\def\arraystretch{1.2}
\begin{tabular}{l|cccc}
& DICE & RAVD & ASSD & MSSD \\ \hline
U-Net& 0.672& 75.092& 66.551& 172.201\\
Deepmedic& 0.905& 10.385& 4.753& 139.120\\
V-Net& 0.828& 19.182& 8.913& 95.328\\
Dense V-networks& 0.902& 8.726& 9.009& 104.886\\
\hline
\rowcolor[rgb]{0.9 0.9 0.9}
Average& 0.827& 28.346& 22.306& 127.884\\
\hline
\rowcolor[rgb]{1 0.8 0.8}
Majority Vote& 0.890& 14.348& 3.341& \fbox{55.303}\\
\rowcolor[rgb]{1 0.8 0.8}
Average& \fbox{0.920}& 7.131& \fbox{3.070}& 74.613\\
\rowcolor[rgb]{1 0.8 0.8}
Product& 0.916& \fbox{6.418}& 3.271& 73.580\\
\rowcolor[rgb]{1 0.8 0.8}
Min-Max& 0.906& 8.799& 3.790& 76.629\\
\end{tabular}
\end{minipage}
\end{table*}

\subsubsection{Ensemble segmenters show less overfitting than individual DMs}
\label{sec:overtrain}
The overfitting can be observed from the tables. Without exception, both for individual DMs and ensembles, the average training values are preferable to the averaged testing values. In some cases the differences are substantial, for example, U-Net gives DICE = 0.904 for the training data and 0.672 for the testing data. Even though the large differences in the metrics values suggest otherwise, further tests revealed that 8 of the 64 differences are not statistically significant at level 0.05. These 8 results are marked with a circle marker in Tables~\ref{tab:res1} and \ref{tab:res3}. 

The results thus far show that while ensembles reach better values of the metrics they are not immune to overfitting. Nonetheless, we will show that, in general, they exhibit less overfitting compared to the individual DMs. To visualise this, we present two tables (\ref{tab:over1} and \ref{tab:over2}), one for each dataset, which contain the overfitting magnitude calculated as training value minus testing value. For DICE, positive difference means that the training value was better. For the other three metrics, negative values indicate that the training value was better because the lower values of these metrics are preferable. The values for each metric are colour-coded. Red colours indicate smaller overfitting while blue colours indicate large overfitting. The blue colour is more present in the top parts of both tables showing that the individual segementers are more prone to that than the ensembles.

\begin{table}
\caption{overfitting magnitude for the CHAOS dataset. Large overfitting corresponds to blue colour and small overfitting, to red colour. Each column (metric) is scaled individually.}\vspace{-2mm}
\label{tab:over1}
\centering
\def\arraystretch{2}
\begin{tabular}{rcccc}
& DICE & RAVD & ASSD & MSSD \\ \hline
U-Net& \cellcolor[rgb]{0.6682 0.6682 1.0000} 0.1238& \cellcolor[rgb]{0.6682 0.6682 1.0000} -40.0423& \cellcolor[rgb]{0.6682 0.6682 1.0000} -10.3499& \cellcolor[rgb]{0.9555 0.9555 1.0000} -49.8649\\
Deepmedic& \cellcolor[rgb]{1.0000 0.7305 0.7305} 0.0329& \cellcolor[rgb]{1.0000 0.6704 0.6704} -1.9436& \cellcolor[rgb]{1.0000 0.9866 0.9866} -5.4651& \cellcolor[rgb]{0.6682 0.6682 1.0000} -74.3951\\
V-Net& \cellcolor[rgb]{1.0000 0.9733 0.9733} 0.0695& \cellcolor[rgb]{1.0000 0.8731 0.8731} -13.6101& \cellcolor[rgb]{1.0000 0.9176 0.9176} -4.4899& \cellcolor[rgb]{0.8218 0.8218 1.0000} -61.2170\\
Dense V-networks& \cellcolor[rgb]{1.0000 0.8196 0.8196} 0.0463& \cellcolor[rgb]{1.0000 0.7171 0.7171} -4.6631& \cellcolor[rgb]{1.0000 0.7550 0.7550} -2.2030& \cellcolor[rgb]{1.0000 0.8731 0.8731} -35.0205\\
\hline
Majority& \cellcolor[rgb]{1.0000 0.6682 0.6682} 0.0237& \cellcolor[rgb]{1.0000 0.6682 0.6682} -1.8338& \cellcolor[rgb]{1.0000 0.6682 0.6682} -0.9726& \cellcolor[rgb]{1.0000 0.6682 0.6682} -17.4741\\
Average& \cellcolor[rgb]{1.0000 0.6882 0.6882} 0.0268& \cellcolor[rgb]{1.0000 0.6860 0.6860} -2.8366& \cellcolor[rgb]{1.0000 0.6927 0.6927} -1.3186& \cellcolor[rgb]{1.0000 0.6860 0.6860} -19.0548\\
Product& \cellcolor[rgb]{1.0000 0.6971 0.6971} 0.0281& \cellcolor[rgb]{1.0000 0.6949 0.6949} -3.3736& \cellcolor[rgb]{1.0000 0.6860 0.6860} -1.2328& \cellcolor[rgb]{1.0000 0.6993 0.6993} -20.1146\\
Min-Max& \cellcolor[rgb]{1.0000 0.7639 0.7639} 0.0381& \cellcolor[rgb]{1.0000 0.7216 0.7216} -4.8866& \cellcolor[rgb]{1.0000 0.7060 0.7060} -1.4997& \cellcolor[rgb]{1.0000 0.7394 0.7394} -23.4931\\
\end{tabular}
\end{table}

\begin{table}
\caption{overfitting magnitude for the 3Dircadb1 dataset. Large overfitting corresponds to blue colour and small overfitting, to red colour. Each column (metric) is scaled individually.}\vspace{-2mm}
\label{tab:over2}
\centering
\def\arraystretch{2}
\begin{tabular}{rcccc}
& DICE & RAVD & ASSD & MSSD \\ \hline
U-Net& \cellcolor[rgb]{0.6682 0.6682 1.0000} 0.2320& \cellcolor[rgb]{0.6682 0.6682 1.0000} -55.7723& \cellcolor[rgb]{0.6682 0.6682 1.0000} -59.4755& \cellcolor[rgb]{0.6860 0.6860 1.0000} -101.4027\\
Deepmedic& \cellcolor[rgb]{1.0000 0.7528 0.7528} 0.0830& \cellcolor[rgb]{1.0000 0.7595 0.7595} -10.1691& \cellcolor[rgb]{1.0000 0.6904 0.6904} -4.3534& \cellcolor[rgb]{0.6682 0.6682 1.0000} -102.9930\\
V-Net& \cellcolor[rgb]{1.0000 0.9777 0.9777} 0.1406& \cellcolor[rgb]{1.0000 0.8352 0.8352} -16.2087& \cellcolor[rgb]{1.0000 0.7327 0.7327} -7.9385& \cellcolor[rgb]{0.9465 0.9465 1.0000} -77.7407\\
Dense V-networks& \cellcolor[rgb]{1.0000 0.7082 0.7082} 0.0714& \cellcolor[rgb]{1.0000 0.7261 0.7261} -7.6161& \cellcolor[rgb]{1.0000 0.7350 0.7350} -8.0785& \cellcolor[rgb]{1.0000 0.7862 0.7862} -53.1059\\
\hline
Majority& \cellcolor[rgb]{1.0000 0.7773 0.7773} 0.0896& \cellcolor[rgb]{1.0000 0.7817 0.7817} -11.9168& \cellcolor[rgb]{1.0000 0.6704 0.6704} -2.7082& \cellcolor[rgb]{1.0000 0.6682 0.6682} -42.3905\\
Average& \cellcolor[rgb]{1.0000 0.6726 0.6726} 0.0626& \cellcolor[rgb]{1.0000 0.6971 0.6971} -5.2867& \cellcolor[rgb]{1.0000 0.6682 0.6682} -2.4555& \cellcolor[rgb]{1.0000 0.8151 0.8151} -55.8543\\
Product& \cellcolor[rgb]{1.0000 0.6682 0.6682} 0.0614& \cellcolor[rgb]{1.0000 0.6682 0.6682} -2.9921& \cellcolor[rgb]{1.0000 0.6682 0.6682} -2.4300& \cellcolor[rgb]{1.0000 0.7817 0.7817} -52.8161\\
Min-Max& \cellcolor[rgb]{1.0000 0.7171 0.7171} 0.0738& \cellcolor[rgb]{1.0000 0.7149 0.7149} -6.7472& \cellcolor[rgb]{1.0000 0.6748 0.6748} -3.0785& \cellcolor[rgb]{1.0000 0.8307 0.8307} -57.2058\\

\end{tabular}
\end{table}

\subsubsection{Ensemble segmenters offer better results than individual DMs}

Tables~\ref{tab:res2} and \ref{tab:res4} show that ensemble values are, in most part, preferable to the values of the individual DMs on all four metrics. We compared the individual DMs and the ensembles on the testing data using our statistical protocol for paired data. Tables~\ref{tab:resdice}-\ref{tab:resmssd} detail the results from the statistical comparisons. The level of significance was set everywhere at 0.05. The tables demonstrate that the ensemble segmenters are better than the individual DMs. 

\begin{table*}
\begin{minipage}{.49\textwidth}
\caption{DICE: Statistical comparison between individual DMs and ensembles. Bullet means that the ensemble wins; circle means that the DM wins; line means that no statistical difference.}\vspace{0mm}
\label{tab:resdice}
\centering
\def\arraystretch{1.3}
\begin{tabular}{rcccc}
\multicolumn{4}{l}{CHAOS dataset}\\
& Majority & Average & Product & Min/Max \\ \hline
U-Net& $\bullet$ & $\bullet$ & $\bullet$ & $\bullet$ \\
Deepmedic& $-$ & $-$ & $-$ & $\circ$ \\
V-Net& $\bullet$ & $\bullet$ & $\bullet$ & $\bullet$ \\
Dense V& $\bullet$ & $\bullet$ & $\bullet$ & $\bullet$ \\
\hline
\\
\multicolumn{4}{l}{3Dircadb1 dataset}\\
& Majority & Average & Product & Min/Max \\ \hline
U-Net& $\bullet$ & $\bullet$ & $\bullet$ & $\bullet$ \\
Deepmedic& $-$ & $-$ & $-$ & $-$ \\
V-Net& $\bullet$ & $\bullet$ & $\bullet$ & $\bullet$ \\
Dense V& $-$ & $-$ & $-$ & $-$ \\
\hline
\end{tabular}
\end{minipage}%
\hspace{5mm}
\begin{minipage}{.49\textwidth}
\caption{RAVD: Statistical comparison between individual DMs and ensembles. Bullet means that the ensemble wins; circle means that the DM wins; line means that no statistical difference.}\vspace{0mm}
\label{tab:resravd}
\centering
\def\arraystretch{1.3}
\begin{tabular}{rcccc}
\multicolumn{4}{l}{CHAOS dataset}\\
& Majority & Average & Product & Min/Max \\ \hline
U-Net& $\bullet$ & $\bullet$ & $\bullet$ & $\bullet$ \\
Deepmedic& $\circ$ & $-$ & $\circ$ & $-$ \\
V-Net& $-$ & $\bullet$ & $-$ & $-$ \\
Dense V& $-$ & $\bullet$ & $-$ & $-$ \\
\hline
\\
\multicolumn{4}{l}{3Dircadb1 dataset}\\
& Majority & Average & Product & Min/Max \\ \hline
U-Net& $\bullet$ & $\bullet$ & $\bullet$ & $\bullet$ \\
Deepmedic& $-$ & $-$ & $-$ & $-$ \\
V-Net& $-$ & $\bullet$ & $\bullet$ & $\bullet$ \\
Dense V& $-$ & $-$ & $-$ & $-$ \\
\hline
\end{tabular}
\end{minipage}%

\begin{minipage}{.49\textwidth}
\vspace{4mm}
\caption{ASSD: Statistical comparison between individual DMs and ensembles. Bullet means that the ensemble wins; circle means that the DM wins; line means that no statistical difference.}\vspace{0mm}
\label{tab:resassd}
\centering
\def\arraystretch{1.3}
\begin{tabular}{rcccc}
\multicolumn{4}{l}{CHAOS dataset}\\
& Majority & Average & Product & Min/Max \\ \hline
U-Net& $\bullet$ & $\bullet$ & $\bullet$ & $\bullet$ \\
Deepmedic& $\bullet$ & $\bullet$ & $\bullet$ & $\bullet$ \\
V-Net& $\bullet$ & $\bullet$ & $\bullet$ & $\bullet$ \\
Dense V& $\bullet$ & $\bullet$ & $\bullet$ & $\bullet$ \\
\hline
\\
\multicolumn{4}{l}{3Dircadb1 dataset}\\
& Majority & Average & Product & Min/Max \\ \hline
U-Net& $\bullet$ & $\bullet$ & $\bullet$ & $\bullet$ \\
Deepmedic& $-$ & $\bullet$ & $\bullet$ & $-$ \\
V-Net& $\bullet$ & $\bullet$ & $\bullet$ & $\bullet$ \\
Dense V& $-$ & $\bullet$ & $-$ & $-$ \\
\hline
\end{tabular}
\end{minipage}%
\hspace{5mm}
\begin{minipage}{.49\textwidth}
\vspace{4mm}
\caption{MSSD: Statistical comparison between individual DMs and ensembles. Bullet means that the ensemble wins; circle means that the DM wins; line means that no statistical difference.}\vspace{0mm}

\label{tab:resmssd}
\centering
\def\arraystretch{1.3}
\begin{tabular}{rcccc}
\multicolumn{4}{l}{CHAOS dataset}\\
& Majority & Average & Product & Min/Max \\ \hline
U-Net& $\bullet$ & $\bullet$ & $\bullet$ & $\bullet$ \\
Deepmedic& $\bullet$ & $\bullet$ & $\bullet$ & $\bullet$ \\
V-Net& $\bullet$ & $\bullet$ & $\bullet$ & $\bullet$ \\
Dense V& $\bullet$ & $\bullet$ & $\bullet$ & $\bullet$ \\
\hline
\\
\multicolumn{4}{l}{3Dircadb1 dataset}\\
& Majority & Average & Product & Min/Max \\ \hline
U-Net& $\bullet$ & $\bullet$ & $\bullet$ & $\bullet$ \\
Deepmedic& $\bullet$ & $\bullet$ & $\bullet$ & $\bullet$ \\
V-Net& $\bullet$ & $-$ & $-$ & $-$ \\
Dense V& $\bullet$ & $\bullet$ & $\bullet$ & $\bullet$ \\
\hline
\end{tabular}
\end{minipage}%
\end{table*}

\begin{figure*}
\begin{minipage}{.50\textwidth}
    \centering
    \includegraphics[width = \textwidth]{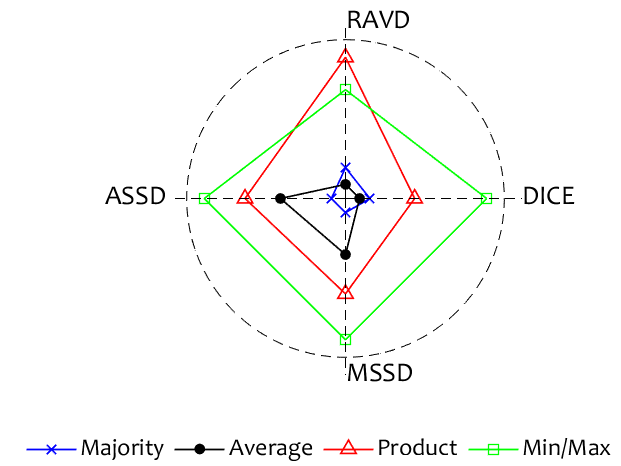}
    \caption{Glyph plot of the four ensemble methods for the CHAOS dataset. The spokes are the four metrics. Small-area ensembles are preferable.}
    \label{fig:glyph1}
\end{minipage}%
\hspace{5mm}
\begin{minipage}{.50\textwidth}
    \centering
    \vspace{-2mm}
    \includegraphics[width = \textwidth]{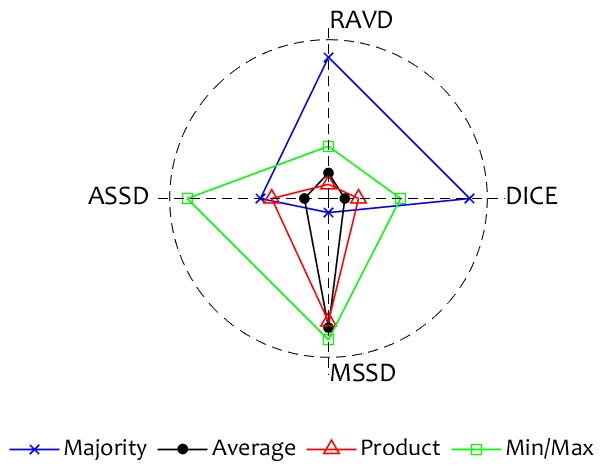}
    \caption{Glyph plot of the four ensemble methods for the 3Dircadb1 dataset. The spokes are the four metrics. Small-area ensembles are preferable.}
    \label{fig:glyph2}
\end{minipage}%
\end{figure*}

Finally, to be able to recommend one of the ensemble models, we present two glyph plots in Figures~\ref{fig:glyph1} and \ref{fig:glyph2}. The plots are based on the averaged testing results for each metric. DICE was reversed so that small values are more desirable. The ensemble scores for each metric were scaled between 0.1 and 1 and plotted on the spokes of the glyph plot. An ideal ensemble would occupy a small square in the middle. The larger the surface of the figure presented by the ensemble, the worse the ensemble in comparison with the rest. The Chaos dataset figure elects the Majority Vote ensemble as the best, closely followed by the Average ensemble. On the other hand, the Majority Vote ensemble occupies a large area in the glyph plot in Fig.~\ref{fig:glyph2}. The Average ensemble is the best for this data set.

Another indication in favour of the Average ensemble is the total number of wins across the data sets, the metrics and the DMs (Tables~\ref{tab:resdice}--\ref{tab:resmssd}). If we add one point for each win, and subtract one point for each loss, the ensembles receive the following scores:
Majority Vote 20, Average 25, Product 21, and Min/Max 20. (All scores are out of possible 2 (data sets) $\times$ 4 (metrics) $\times$ 4 (DMs) = 32 points.) This gives us ground to recommend the Average combiner for future use with our vanilla-style ensemble of DMs for liver segmentation.

\section{Conclusion}
\label{sec:concl}

Intrigued by the success of DMs in medical segmentation, we set off to explore the potential of an ensemble of DM segmenters which is composed of publicly available, state-of-the-art DMs, and which does not require profound expertise in deep learning or ensemble methods from the user. The first problem we encountered was that the individual DMs (trained with their default structure, parameters and training setting) are prone to overfitting. Our experimental results confirmed that, and also revealed that the propose vanilla-style ensemble is less affected by overfitting. Using two publicly available datasets, we demonstrated that a simple ensemble of off-the-shelf deep learning models outperforms the individual ensemble members. The overall message of this study is that it is possible to achieve results in liver segmentation from CT images with minimal programmatic effort, using state-of-the-art DMs as black boxes and basic classifier ensemble rules. Besides, since data/problem-specific design and parameter tuning are not required, the ensemble methods shown in this study can be offered as a solution to the repeatability and reproducibility problems widely seen in deep learning studies. Out of the four combination rule we examined, the average (or sum) combiner achieved the best result.

In addition to the chosen simple combiners, we experimented with several trained combiners: weighted majority vote, Na\"ive Bayes and Behaviour Knowledge Space (BKS)~\cite{Kuncheva2014}. The results were on a par with the simple combiners. This reinforces our message that for small data sizes as the currently available annotated data for liver segmentation, overfitting is a major issue.

By and large, the combination methods recommended in the literature have been evaluated on {\em classification accuracy}. Here we note that our evaluation hinges on four different metrics. They are not straightforwardly related to classification accuracy. This suggests that developing bespoke training protocols for the DMs as well as more suitable combination strategies could be a good way forward.

We note that the success of the combiners which rely on continuous-valued outputs (Average, Product and Min/Max) critically depends on the calibration of the output of the segmenters. Traditionally, ensembles are constructed of the same base classifier (segmenter here) with different training data, which practically eliminates the problem of calibration. However, if different models are used, as in this study, it is vital to ensure that the probability map calibration is suitable. For example, if one segmenter is very ``certain'' in its decision and always gives values close to 0 and close to 1, this segmenter will have a heavier vote. The segmenter may be arbitrarily accurate and may not deserve the advantage over the rest of the segmenters. In this study, we were striving for simplicity and did not calibrate further the four DMs outputs. On the other hand, Dede et al.~\cite{Dede19} observe that heterogeneous ensembles of DMs fare better than homogeneous ones, which the authors attribute to the importance of diversity offered by different DM models. Thus, it may pay off to devise and include a calibration pre-processing step in the ensemble pipeline.

\section*{Acknowledgement}
This work is supported by Scientific and Technological Research Council of Turkey (TUBITAK) with 2214 International Doctoral Research Fellowship Programme and TUBITAK ARDEB-EEEAG under grant number 116E133. The authors gratefully acknowledge the support of NVIDIA Corporation with the donation of the Titan Xp Pascal GPU used for this research.

\IEEEtriggeratref{48}
\bibliographystyle{IEEEtranTKDE}
\bibliography{LiverSegmentationEnsemble_TKDE}

\enlargethispage{-7cm}
\begin{IEEEbiographynophoto}{Ali Emre Kavur} received the BSc degree from Izmir Institute of Technology, Izmir, Turkey, in 2006, and the MSc degrees from Dokuz Eylül University, Izmir, Turkey in 2011 and from Katip Celebi University, Izmir, Turkey in 2018. Currently, he is a Ph.D. candidate at Dokuz Eylül University, Izmir, Turkey. During his Ph.D., he made an academic visit to Dr. Ludmila I. Kuncheva's lab at the School of Computer Science and Electronic Engineering, Bangor University, Bangor, UK. His main research interests are medical image processing, organ segmentation, classifier ensemble studies, and grand challenges in biomedical image analysis. \vspace*{-4\baselineskip}
\end{IEEEbiographynophoto}

\begin{IEEEbiographynophoto}{Ludmila I. Kuncheva} received the M.Sc. degree from the Technical University of Sofia, Bulgaria, in 1982, and the Ph.D. degree from the Bulgarian Academy of Sciences, Sofia, Bulgaria, in 1987. She is currently a Professor at the School of Computer Science and Electronic Engineering, Bangor University, Bangor, UK. She has published two books and above 200 scientific papers. Her current research interests include pattern recognition and classification, machine learning, and classifier ensembles. Dr. Kuncheva is a Fellow of the International Association for Pattern Recognition. She is also the recipient of two Best Paper Awards (IEEE Transactions on SMC, 2002, and IEEE Transactions on Fuzzy Systems, 2006).
\vspace*{-4\baselineskip}
\end{IEEEbiographynophoto}

\begin{IEEEbiographynophoto}{M. Alper Selver} received the BSc degree from Gazi University, Ankara, Turkey, in 2002, and the MSc and PhD degrees from Dokuz Eylül University, Izmir, Turkey, in 2005 and 2010, respectively, all in electrical and electronics engineering. During his graduate studies, he has studied in Medical Informatics Laboratory at FH-Aachen Abt. Juelich, Germany and Heffner Biomedical Imaging Laboratory at Columbia University, New York, USA. Since 2011, he has been working as an associate professor at the DEU EEE. His main research interests include the field of radiological image processing- visualisation, hierarchical, and multiscale classification strategies for biomedical applications and software development for the use of the developed techniques in the clinic.
\end{IEEEbiographynophoto}

\end{document}